\documentstyle[preprint,epsfig,eqsecnum,aps,prd,floats]{revtex}
\global\firstfigfalse
\global\firsttabfalse
\tightenlines
\begin{document}
\draft
\preprint{UM-P-018-2002}
\title{Approximations to the QED Fermion Green's Function in a
Constant External Field}
\author{S.K.J. Daniel and B.H.J. McKellar}
\address{
School of Physics,
University of Melbourne,
Parkville, Victoria 3052,
Australia}
\date{10 October 2002}
\maketitle

\newcommand{\eqn}[1]{(\ref{#1})}
\newcommand{\ds}{\displaystyle}
\newcommand{\ts}{\textstyle}
\newcommand{\be}{\begin{equation}}
\newcommand{\ee}{\end{equation}}
\newcommand{\ba}{\begin{eqnarray}}
\newcommand{\ea}{\end{eqnarray}}
\newcommand{\bi}{\bibitem}
\newcommand{\intl}{\int\limits}
\newcommand{\suml}{\sum\limits}
\newcommand{\prodl}{\prod\limits}
\newcommand{\Ln}{\Lambda_{N}}
\newcommand{\sn}{\sin\left({N\pi s'\over s}\right)}
\newcommand{\cn}{\cos\left({N\pi s'\over s}\right)}
\newcommand{\Lone}{\Lambda_{1}}
\newcommand{\sone}{\sin\left({\pi s'\over s}\right)}
\newcommand{\cone}{\cos\left({\pi s'\over s}\right)}
\newcommand{\Ltwon}{\Lambda_{2n-1}}
\newcommand{\stwon}{\sin\left({(2n-1)\pi s'\over s}\right)}
\newcommand{\ctwon}{\cos\left({(2n-1)\pi s'\over s}\right)}
\newcommand{\Det}{\hbox{Det}}
\newcommand{\ints}{\int_0^s ds'\,}
\newcommand{\zero}{\left[\matrix{0&0\cr 0&0\cr}\right]}
\newcommand{\ann}{\left[\matrix{A^{(N,N)}\cr{}\cr}\right]}
\newcommand{\sech}{\hbox{sech}}

\begin{abstract}
An exact representation of the causal QED fermion Green's function, in
an arbritrary external electromagnetic field, derived in \cite{Fried
McKellar paper}, and which naturally allows for non-perturbative
approximations, is here used to calculate non-perturbative
approximations to the Green's function in the simple case of a
constant external field.  Schwinger's famous exact result is obtained
as the limit as the order of the approximation approaches infinity.
\end{abstract}

\section{Introduction}

An exact representation for the causal QED fermion Green's function,
$G_{c}(x,y\, |A)$, in an arbitrary external field, 
was derived by Fried et al.$\!$\cite{Fried McKellar paper} in such a way 
as to obtain an exact representation which naturally allows for
non-perturbative approximations.  The rather intimidating exact
representation is 

\ba
G_{c}(x,y\, |A) =
&i&\!\int_{0}^{\infty}ds\,e^{-ism^2}\int{d^{4}p\over(2\pi)^4}
e^{ip\cdot(x-y)}\int{d^{4}z d^{4}P\over(2\pi)^{4}}
e^{iP\cdot(z-y)+i{s\over4}P^{2}}
\nonumber\\
&\times& 
{\prod_{N=1}^{\infty}}\!'{(-i)^{2}\over(2\pi)^{4}}\int d^{4}P_{N}d^{4}Q_{N}
e^{{i \over 2}(P_{N}^{2}+Q_{N}^{2})}
\nonumber\\
&\times&
e^{-i\ints[p-\Omega(s')]^{2}}\, 
e^{-g\ints{\partial\over\partial z_{\mu}}A_{\mu}\left
(\zeta(s')-2\int_{0}^{s'}\Omega\right)}
\nonumber\\
&\times&
\{m-i\gamma\cdot[p-\Omega(s)]\}
\,{\left( e^{\, g\ints\sigma\cdot F\left(\zeta(s') 
-2\int_{0}^{s'}\Omega\right)}\right)}_+, \label{g1}
\ea
where $\Omega(s')$ is the solution to the ``map''

\be
\Omega(s')=gA\left(\zeta(s')-2\int_{0}^{s'}\Omega\right), \label{omega}
\ee
and $\zeta(s')$ is given by
\be
\zeta(s')=z+s'(2p+P)-{2\sqrt{s}\over\pi}\sum_{N=1}^{\infty}\!'{1\over N}
\left[P_{N}\cos\left({N\pi s'\over s}\right) 
+Q_{N}\sin\left({N\pi s'\over s}\right)\right]. \label{zeta}
\ee
${{\prod_{N=1}^{\infty}}\!\!\!\!'}$$\;$ and
${\sum_{N=1}^{\infty}}\!\!\!\!'$$\;$ represent the product and sum
respectively over all odd natural numbers.  It is easy to see that we
may approximate the Green's function non-perturbatively by retaining a
finite number of integrations, in particular, for the $n$th
approximation $G_{c}^{(n)}(x,y\,| A)$, where $n=0,1,2,\ldots$, we
retain $2n+3$ integrations.  This means that for $n\not=0$, the odd
number $N$ takes the values $N=1,3,\ldots,2n-1$, while for the zeroth
approximation all $N$-dependence is neglected.  The exact result is of
course recovered in the limit $n\to\infty$, that is, $G_{c}(x,y\,| A)=
\lim_{n\to\infty}G_{c}^{(n)}(x,y\,| A)$.

It is well known that there is an exact expression, first obtained by
Schwinger \cite{Schwinger}, for the 
fermion Green's function in the simple case 
of a constant, but otherwise arbitrary, external field, which (in an arbitrary
gauge) reads

\ba
G_{c}(x,y\,| A)=&{}&\!\!\!\!\!\!\Phi(x,y\,| A){1\over(4\pi)^2}\int_0^{\infty}
{ds\over s^2}\,e^{-ism^{2}}e^{gs\sigma\cdot F}
\left(\det{\sinh gFs\over gFs}\right)^{-{1\over2}}
\nonumber\\
&\times&e^{{i\over4}(x-y)gF\coth gFs(x-y)}
\left[m-{i\over2}\gamma\cdot(gF\coth gFs+gF)(x-y)\right], 
\label{schwinger}
\ea
where the holonomy factor $\Phi(x,y\,| A)$ is

\be
\Phi(x,y\,| A)=e^{ig\int_y^x d\xi_{\mu}\,[A_{\mu}(\xi)+{1\over2}F
(\xi-y)]},
\ee
and carries the complete gauge-dependence of the Green's function.
It is convenient to employ matrix 
notation, in which we regard the 
field strength tensor as a constant, antisymmetric $4\times4$ matrix $F$.

It is obvious that Schwinger's result must somehow be contained as a
special case of the exact representation \eqn{g1}-\eqn{zeta}.  Furthermore, the
Fradkin representation \cite{Fradkin},\cite{Fried book}, from which
the above representation was derived, almost trivially yields
Schwinger's result.  This leads us to expect that the latter may be
extracted, analytically, from the above representation, and here we
demonstrate that this is indeed the case.

We proceed by evaluating that $n$th non-perturbative approximation to
the Green's function in a constant field as given by 
\eqn{g1}-\eqn{zeta} 
Schwinger's result is then recovered in the limit as the order of this
approximation approaches infinity.  That we may carry out this
programme relies on two things: for the case of constant $F$ the
ordered exponential of \eqn{g1} becomes an ordinary exponential, and
the $2n+3$ integrations we must perform are all Gaussian.  Indeed a
large part of the evaluation of the $n$th approximation is an
extension of our matrix notation to account for this latter fact. 
$2n+3$ Gaussian integrals may be expressed as one Gaussian integral
over a $2n+3$-dimensional space.  The integration is then, with the
appropriate notation, essentially trivial.  In Section II we make the
change of notation, perform the integration, and express the $n$th
approximation in terms of the two resulting, order $2n+3$,
determinants.  In Section III we evaluate the determinants and take
the limit $n\to\infty$ to recover Schwinger's result.

We emphasise that the result when stopped at a finite $n$ (except $n = 
0$) is essentially non-perturbative, in that it is \emph{not} a 
polynomial in $g$.  The result of \eqn{g1}-\eqn{zeta} is thus, as $n$ is 
increased, a systematically improving, but non-perturbative, method for 
calculating the Green's function in an external field.

\section{The Integration}
We must first choose a gauge to work in.  It shall be convenient to
work exclusively in the Schwinger-Fock ($SF$) gauge

\be
A_{\mu}^{SF}(z)=-{1\over2}F_{\mu\nu}(z-y)_{\nu},
\ee
the initial motivation for which is that we may forget about the holonomy factor, 
which reduces to 1. An immediate conseqence is that 
${\partial\over\partial z_{\mu}}A_{\mu}^{SF}(z)=0$, due to the 
antisymmetry of $F$; the factor in \eqn{g1} containing this term in the exponent also 
reduces to 1.  With the above simplifications, the $n$th approximation to the fermion
Green's function in a constant field, in the $SF$ gauge, as given by 
\eqn{g1}, is thus
\ba
G_{c}^{(n)}(x,y\,| A^{SF})=
&&\!\!\!\!\!\!\int_{0}^{\infty}ds\,e^{-ism^{2}}e^{gs\sigma\cdot F}
{i\over(2\pi)^8}{(-i)^{2n}\over(2\pi)^{4n}}\int dp\,dz\,dP\,
{\prodl_{N=1}^{2n-1}}\!'dP_{N}\,dQ_{N}
\nonumber\\
&\times&e^{ip\cdot(x-y)+iP\cdot(z-y)+i{s\over4}P^{2}
+\suml_{N=1}^{2n-1}\!\!'{i\over2}(P_{N}^2+Q_{N}^2)}
\nonumber\\
&\times&e^{-i\ints[p-\Omega^{(n)}(s')]^2}
\{m-i\gamma\cdot[p-\Omega^{(n)}(s)]\}, \label{g2}
\ea
where $\Omega^{(n)}(s')$ is
the solution to \eqn{omega}, \eqn{zeta}, but with the sum in \eqn{zeta}
terminating at $N=2n-1$, and in the spirit of using matrix notation we
have used $dp$ instead of $d^4p$ and so on.  In the $SF$ gauge,
\eqn{omega}, \eqn{zeta} yield a simple integral equation for $\Omega^{(n)}(s')$:
\ba
&&\Omega^{(n)}(s')-gF\int_{0}^{s'}\Omega^{(n)}
\nonumber\\
=&&-{1\over2}gF\left\{z-y+s'(2p+P)
-{2\sqrt{s}\over\pi}
\sum_{N=1}^{2n-1}\!\!'{1\over N}\left[P_{N}\cn+Q_{N}\sn\right]\right\}.
\ea
The equivalent differential equation plus boundary condition may be solved 
with only elementary integrals.  It is the combination $p-\Omega^{(n)}(s')$
which appears in \eqn{g2}, we find
\ba
p-\Omega^{(n)}(s')=
&e&^{\!\!\!gFs'}p+{1\over2}gFe^{gFs'}(z-y)+{1\over2}(e^{gFs'}-1)P
\nonumber\\
&-&{1\over\sqrt{s}}\sum_{N=1}^{2n-1}\!\!' 
{1\over1+\Ln^2}\left[
\sn+\Ln\cn+{1\over\Ln}e^{gFs'}\right]P_{N}
\nonumber\\
&-&{1\over\sqrt{s}}\sum_{N=1}^{2n-1}\!\!'{1\over1+\Ln^2}\left[
-\cn+\Ln\sn+e^{gFs'}\right]Q_{N}, \label{pomega}
\ea
where $\Ln={N\pi\over gFs}$.  Our choice of gauge allows us to make the 
change of variable $z-y\to z$, after which $p-\Omega^{(n)}(s')$ is 
independent of $y$, so that the only $x$ and $y$ dependence in the 
exponent of \eqn{g2} appears in the term $ip\cdot(x-y)$.  

If we imagine substituting \eqn{pomega} into \eqn{g2}, we recognise
that all of the terms in the exponent of the integrand, save the term
$ip\cdot(x-y)$, are able to be expressed in the form
${i\over2}X_i^TA_{ij}^{(n)}X_j$, where the $X_i$ are the $2n+3$
4-vector variables we must integrate over, the $A_{ij}^{(n)}$ are some
matrix functions of $F$, and the ${i\over2}$ is a convenient
normalisation factor.  This suggests that we extend our matrix
notation, and write this part of the exponent of \eqn{g2} as
${i\over2}X^TA^{(n)}X$, where $X$ is a $(2n+3)\times1$ column vector
of 4-vector variables, and $A^{(n)}$ is a $(2n+3)\times(2n+3)$
symmetric matrix with matrix elements $A_{ij}^{(n)}$.  Now
$dpdzdP\prodl_{N=1}^{2n-1}\!\!\!'dP_NdQ_N=d^{2n+3}X$.  Let us define
the column vector $X$ such that

\be
X^{T}=[p\;\;z\;\;P\;\;P_{1}\;\;Q_{1}\;\;\ldots\;\;P_{N}\;\;Q_{N}\;\;\ldots\;\;
P_{2n-1}\;\;Q_{2n-1}].
\ee
The term $ip\cdot(x-y)$ is included by introducing 

\be 
B^{(n)T}=[(x-y)\;\; 0\;\; 0\;\;\ldots\;\; 0], 
\ee
whence $e^{ip\cdot(x-y)}=e^{iB^{(n)T}X}$.  A construction for the matrix
$A^{(n)}$ is obtained by defining

\be p-\Omega^{(n)}(s')=C^{(n)T}(s')X, \ee
so that
$e^{-i\ints[p-\Omega^{(n)}(s')]^2}=e^{-iX^T\ints
C^{(n)}(s')C^{(n)T}(s')X}$.  The elements of the row vector
$C^{(n)T}(s')$ are read straight from \eqn{pomega}.  The transpose of
the row vector
 is

\be
C^{(n)}(s')=
\left[
\matrix{e^{-gFs'} \cr
-{1\over 2}gFe^{-gFs'} \cr
{1\over 2}(e^{-gFs'}-1) \cr
{1\over\sqrt{s}}{1\over 1+\Lone^2}
\left[-\sone+\Lone\cone+{1\over\Lone}e^{-gFs'}\right] \cr
\vspace{-2.0mm}{1\over\sqrt{s}}{1\over 1+\Lone^2}
\left[\cone+\Lone\sone-e^{-gFs'}\right] \cr
\vspace{-2.0mm}\vdots\cr
{1\over\sqrt{s}}{1\over1+\Ln^2}\left[-\sn+\Ln\cn+{1\over\Ln}e^{-gFs'}\right]\cr
\vspace{-2.0mm}{1\over\sqrt{s}}{1\over 1+\Ln^2}\left[\cn+\Ln\sn-e^{-gFs'}\right] \cr
\vdots\cr}
\right], \label{colveccn}
\ee
where we have used $F^T=-F$ $(\Ln^T=-\Ln)$, and omitted the last two $N=2n-1$ 
elements for brevity. We also write 
$e^{iP\cdot z+i{s\over4}P^{2}+\suml_{N=1}^{2n-1}\!'\,{i\over2}
(P_{N}^2+Q_{N}^2)}=e^{{i\over2}X^TD^{(n)}X}$, which defines the matrix $D^{(n)}$:

\be
D^{(n)}=
\left[\matrix{
\vspace{-2.0mm}0&0&0&0&0&\cdots&0&0\cr
\vspace{-2.0mm}0&0&1&0&0&\cdots&0&0\cr
\vspace{-2.0mm}0&1&{s\over2}&0&0&\cdots&0&0\cr
\vspace{-2.0mm}0&0&0&1&0&\cdots&0&0\cr
\vspace{-2.0mm}0&0&0&0&1&\cdots&0&0\cr
\vspace{-2.0mm}\vdots&\vdots&\vdots&\vdots&\vdots&\ddots&\vdots&\vdots\cr
\vspace{-2.0mm}0&0&0&0&0&\cdots&1&0\cr
0&0&0&0&0&\cdots&0&1\cr}
\right].
\ee
Our change of notation complete, \eqn{g2} and \eqn{pomega} become 

\ba
G_{c}^{(n)}(x,y\,| A^{SF})=&&\!\!\!\!\!\!\int_0^{\infty}ds\,e^{-ism^2}e^{gs\sigma\cdot F}
{i\over(2\pi)^8}{(-i)^{2n}\over(2\pi)^{4n}}
\nonumber\\
&\times&\int d^{2n+3}X\,e^{{i\over2}
X^TA^{(n)}X+iB^{(n)T}X}[m-i\gamma\cdot C^{(n)T}(s)X],
\ea 
where 

\be
A^{(n)}=-2\ints C^{(n)}(s')C^{(n)T}(s')+D^{(n)}. \label{an}
\ee 

All that we have done is to re-express, in the usual way, the product of $2n+3$ Gaussian integrals 
as one $(2n+3)$-dimensional Gaussian integral.  After a change of variable
$X\to X-A^{(n)-1}B^{(n)}$, the integration is trivial, we obtain

\ba
G_{c}^{(n)}(x,y\,| A^{SF})=
&{1\over4\pi^2}&\int_0^{\infty}ds\,e^{-ism^2}
e^{gs\sigma\cdot F}(\Det\,A^{(n)})^{-{1\over2}}e^{-{i\over2}
B^{(n)T}A^{(n)-1}B^{(n)}}
\nonumber\\
&\times&
[m+i\gamma\cdot C^{(n)T}(s)A^{(n)-1}B^{(n)}],
\ea
where $\Det\,A^{(n)}$ is the determinant of $A^{(n)}$.

The matrix $A^{(n)}$is a $4(2n+3)\times4(2n+3)$ matrix which is naturally partitioned
into the $(2n+3)\times(2n+3)$ $4\times4$ matrices 
$A_{ij}^{(n)}=-2\ints C_i^{(n)}(s')C_j^{(n)T}(s')+D_{ij}^{(n)}$.  The 
$A_{ij}^{(n)}$, as (matrix) functions of $F$ only, commute, and are thus referred to 
as the ``elements'' of $A^{(n)}$.  Since $A^{(n)}$ is symmetric, we have the 
relations $A_{ii}^{(n)}=A_{ii}^{(n)T}$ and  $A_{ji}^{(n)}=A_{ij}^{(n)T}$.
If we form a determinant using the elements $A_{ij}^{(n)}$, the result will be
a $4\times4$ matrix, also some function of $F$, which we denote $\det\,A^{(n)}$.
It is easy to convince oneself that the determinant of $A^{(n)}$, which may be 
partitioned in this way, is the determinant of the matrix we have called 
$\det\,A^{(n)}$, that is, $\Det\,A^{(n)}=\det(\det\,A^{(n)})$.

Recalling that $B_i^{(n)}=(x-y)\delta_{i1}$, and with the notation discussed above, 
we may write

\be
C^{(n)T}(s)A^{(n)-1}B^{(n)}={\det A_1^{(n)}\over\det A^{(n)}}(x-y),
\ee 
where $A_1^{(n)}$ is the matrix obtained by replacing the first row of 
$A^{(n)}$ by $C^{(n)T}(s)$. Similarly,

\be
B^{(n)T}A^{(n)-1}B^{(n)}=(x-y){\det\tilde A^{(n)}\over\det A^{(n)}}(x-y),
\ee
where $\det\tilde A^{(n)}$ is the $(1,1)$ cofactor of $A^{(n)}$, that is, 
$\tilde A^{(n)}$ is the matrix obtained by deleting the first row and the first 
column of $A^{(n)}$.  In fact it will be shown in the next section that 
${\det\tilde A^{(n)}\over\det A^{(n)}}$ is the part of 
${\det A_1^{(n)}\over\det A^{(n)}}$ symmetric with respect to the interchange
of space-time indices, for all $n$, and denoted by a superscript $S$.  
The $n$th approximation to the Green's function, 
in terms of the two determinants $\det A^{(n)}$ and $\det A_1^{(n)}$, is then

\ba
G_{c}^{(n)}(x,y\,| A^{SF})=&{1\over 4\pi^2}&\int_0^{\infty}ds\,e^{-ism^2}
e^{gs\sigma\cdot F}(\Det\,A^{(n)})^{-{1\over 2}}e^{-{i\over 2}
(x-y)\left({\det A_1^{(n)}\over\det A^{(n)}}\right)^S(x-y)}
\nonumber\\
&\times&\left[m+i\gamma\cdot{\det A_1^{(n)}\over\det A^{(n)}}(x-y)\right].
\label{gnth}
\ea 

Comparing \eqn{gnth} with Schwinger's result \eqn{schwinger}, and letting 
$\det A=\lim_{n\to\infty}\det A^{(n)}$ and 
$\det A_1=\lim_{n\to\infty}\det A_1^{(n)}$, it is necessary that 

\be
\det A=2s{\sinh gFs\over gFs}, \label{deta}
\ee
so that $(\Det A)^{-{1\over2}}={1\over4s^2}(\det{\sinh gFs\over gFs})^{-{1\over2}}$, and

\be
\det A_1=-e^{gFs}, \label{deta1}
\ee
so that the quotient ${\det A_1\over\det A}=-{1\over2}(gF\coth gFs+gF)$.  Note that the 
symmetric part of ${\det A_1\over\det A}$ is 
$({\det A_1\over\det A})^S=-{1\over2}gF\coth gFs$, we mentioned above that this 
relation holds for all values of $n$.  Thus $\det A^{(n)}$ and 
$\det A_1^{(n)}$ are yet to be determined, $n$th, non-perturbative approximations to 
\eqn{deta} and \eqn{deta1} respectively.  The quotient 
${\det A_1^{(n)}\over\det A^{(n)}}$ provides a non-perturbative approximation to 
$-{1\over2}(gF\coth gFs+gF)$, the symmetric part of the former a non-perturbative 
approximation to the symmetric part of the latter.  In the next section we calculate
exact expressions for these non-perturbative approximations, and show that we can 
obtain \eqn{deta} and \eqn{deta1}, and thus Schwinger's result, in the limit $n\to\infty$. 

\section{The Determinants}
An obvious but important fact is that we only need to find the determinants of the 
matrices $A^{(n)}$ and $A_1^{(n)}$, not the matrices themselves.  This means that we
can simplify $C^{(n)}(s')$, its transpose, and $D^{(n)}$, with any row and column 
operations which do not alter the determinant, before using \eqn{deta1} to find $A^{(n)}$.
We now find the reduced form of $A^{(n)}$, from this it will be easy to obtain the 
reduced form of $A_1^{(n)}$ and of $\tilde A^{(n)}$.  We perform the following 
sets of row operations on the column vector $C^{(n)}(s')$: 
use the first element to remove those
terms proportional to $e^{-gFs'}$ from all other elements, noting that the second
element requires the row operation row$_2$ $\to$ row$_2 + {1\over 2}gF$row$_1$; then
row$_{N+3}$ $\to$ row$_{N+3}-\Ln$row$_{N+4}$; then 
row$_{N+4}\to$ row$_{N+4}+{\Ln -{1\over\Ln}\over 1+\Ln^2}$row$_{N+3}$; the last two sets
for all rows $N=1,3,\ldots,2n-1$.  $C^{(n)}(s')$ becomes

\be
C^{(n)}(s')\sim\left[
\matrix{ e^{-gFs'}\cr
0\cr
-{1\over 2}\cr
-{1\over\sqrt{s}}\sone \cr
\vspace{-2.0mm}{1\over\sqrt{s}}{1\over 1+\Lone^2}
\left[\cone+{1\over\Lone}\sone\right] \cr
\vspace{-2.0mm}\vdots\cr
-{1\over\sqrt{s}}\sn\cr
\vspace{-2.0mm}{1\over\sqrt{s}}{1\over 1+\Ln^2}\left[\cn+{1\over\Ln}\sn\right] \cr
\vdots\cr}
\right]. \label{veccn}
\ee
The row operations are performed on $C^{(n)}(s')$ and $D^{(n)}$.  To keep things 
symmetric we perform the transposed operations on $C^{(n)T}(s')$ and $D^{(n)}$.  
The integrals required by \eqn{deta1} are elementary and we easily obtain the reduced
form of the matrix $A^{(n)}$:

\be
A^{(n)}\sim\left[\matrix{
-2s&0&-{1\over gF}(e^{-gFs}-1)&x_1^T&0&\cdots&x_N^T&0&\cdots\cr
0&0&1&0&0&\cdots&0&0&\cdots\cr
{1\over gF}(e^{gFs}-1)&1&0&0&0&\cdots&0&0&\cdots\cr
x_1&0&0&-\Lone^2&-{3\Lone\over1+\Lone^2}&\cdots&0&0&\cdots\cr
\vspace{-2.0mm}0&0&0&{3\Lone\over1+\Lone^2}&{5-\Lone^2\over(1+\Lone^2)^2}&\cdots&0
&0&\cdots\cr
\vspace{-2.0mm}\vdots&\vdots&\vdots&\vdots&\vdots&\ddots&\vdots&\vdots&{}\cr
x_N&0&0&0&0&\cdots&-\Ln^2&-{3\Ln\over1+\Ln^2}&\cdots\cr
\vspace{-2.0mm}0&0&0&0&0&\cdots&{3\Ln\over1+\Ln^2}&{5-\Ln^2\over(1+\Ln^2)^2}&\cdots\cr
\vdots&\vdots&\vdots&\vdots&\vdots&{}&\vdots&\vdots&\ddots\cr}\right], 
\label{matan}
\ee
where $x_N={2\over\sqrt{s}}{\Ln\over1+\Ln^2}{1\over gF}(1+e^{gFs})$ and we have
further used the $(2,3)$ and $(3,2)$ elements of the reduced $A^{(n)}$ to eliminate
those elements to the right of the former and below the latter.
After trivially expanding the determinant along the second row and down the second
column we are left with the determinant of a $(2n+1)\times(2n+1)$ bordered matrix. 
The first set of row and column operations and the orthogonality of the sine and 
cosine functions have ensured the matrix is block diagonal, the other sets of operations 
that every second element of the border is zero.  Expanding along the first row and 
down the first column we obtain the following expression for the $n$th non-perturbative
approximation to \eqn{deta}:

\be
\det A^{(n)}=2s\left(\prod_{N=1}^{2n-1}\!'\alpha_N\right)
\left[1+{8\over\lambda^2}\cosh^2\left({\lambda\over2}\right)
\left(\sum_{N=1}^{2n-1}\!'\beta_N\right)\right], \label{detan}
\ee 
where

\be
\alpha_N=
{1+{4\lambda^2\over N^2\pi^2}\over\left(1+{\lambda^2\over N^2\pi^2}\right)^2}\;,\; 
\beta_N={5-{N^2\pi^2\over\lambda^2}\over
\left(4+{N^2\pi^2\over\lambda^2}\right)\left(1+{N^2\pi^2\over\lambda^2}\right)^2}\;,
\ee
and we have written the approximation in terms of $\lambda=gFs$, 
$(\Ln={N\pi\over\lambda})$.
That \eqn{detan} is an approximation to \eqn{deta} can be seen with the help of the 
relations \cite{Smithsonian}

\be
\cosh x=\prod_{N=1}^{\infty}\!'\left(1+{4x^2\over N^2\pi^2}\right),\; 
\quad
{x\over2}\tanh x=\sum_{N=1}^{\infty}\!'{1\over 1+{N^2\pi^2\over 4x^2}}\;,
\label{prodsum}
\ee
whence

\be
\prod_{N=1}^{\infty}\!'\alpha_N={\cosh\lambda\over\cosh^2\left({\lambda\over2}\right)}\;,\;
\sum_{N=1}^{\infty}\!'\beta_N={\lambda\over8}\tanh\lambda
-{\lambda^2\over8}\sech^2\left({\lambda\over2}\right).
\ee
The nature of the approximation is now evident. The function 
${\sinh\lambda\over\lambda}$ is rewritten as 
${\cosh\lambda\over\cosh^2\left({\lambda\over2}\right)}
\left\{1+{8\over\lambda^2}\cosh^2\left({\lambda\over2}\right)
\left[{\lambda\over8}\tanh\lambda-{\lambda^2\over8}\sech^2\left({\lambda\over2}\right)
\right]\right\}$, the foremost factor is expressed exactly as the infinite product of the 
$\alpha_N$s, and the expression in the square brackets as the infinite sum of the 
$\beta_N$s.  The $n$th approximation is then defined by including the first $n$ terms
in the product and in the sum.

Note that $\det A^{(n)}$ is an even function of $\lambda$ (of $F$), and hence 
symmetric, for all $n$.  This is desirable since the exact \eqn{deta} is symmetric.
The zeroth approximation, in which \eqn{deta} is approximated
by $\det A^{(0)}=2s$, is the only perturbative result, of order $(gFs)^0$.  Every approximation order
greater than zero contains all (natural number) powers of $gFs$.

The matrix \eqn{matan} shall be our starting point for finding $\det A_1^{(n)}$.
Before replacing the first row of \eqn{matan} with the column-reduced form of 
$C^{(n)T}(s)$ (the transpose of \eqn{veccn} with $s'=s$), we must reinstate the 
second row via row$_2\to$ row$_2-{1\over 2}gF$row$_1$.  This procedure is 
valid since the second element of $C^{(n)}(s')$ in \eqn{colveccn} is proportional to the
first element, and we could have used the second element to eliminate the
$e^{-gFs'}$ terms instead of the first element.  After thus reinstating the second row
and replacing the first row with the column-reduced $C^{(n)T}(s)$, we obtain the 
reduced form of the (neither symmetric nor antisymmetric) matrix $A_1^{(n)}$:

\be
\left[\matrix{
e^{gFs}&0&-{1\over2}&0&-{1\over\sqrt{s}}{1\over{1+\Lone^2}}&\cdots&0&-{1\over\sqrt{s}}{1\over{1+\Ln^2}}&\cdots\cr
gFs&0&{1\over2}(e^{-gFs}+1)&-{1\over2}gFx_1^T&0&\cdots&-{1\over2}gFx_N^T&0&\cdots\cr
{1\over gF}(e^{gFs}-1)&1&0&0&0&\cdots&0&0&\cdots\cr
x_1&0&0&-\Lone^2&-{3\Lone\over1+\Lone^2}&\cdots&0&0&\cdots\cr
\vspace{-2.0mm}0&0&0&{3\Lone\over1+\Lone^2}&{5-\Lone^2\over(1+\Lone^2)^2}&\cdots&0
&0&\cdots\cr
\vspace{-2.0mm}\vdots&\vdots&\vdots&\vdots&\vdots&\ddots&\vdots&\vdots&{}\cr
x_N&0&0&0&0&\cdots&-\Ln^2&-{3\Ln\over1+\Ln^2}&\cdots\cr
\vspace{-2.0mm}0&0&0&0&0&\cdots&{3\Ln\over1+\Ln^2}&{5-\Ln^2\over(1+\Ln^2)^2}&\cdots\cr
\vdots&\vdots&\vdots&\vdots&\vdots&{}&\vdots&\vdots&\ddots\cr}\right],
\ee 
We may now expand
the determinant to obtain the following $n$th approximation to \eqn{deta1}:

\be
\det A_1^{(n)}=-{1\over 2}\left(\prod_{N=1}^{2n-1}\!'\alpha_N\right)
\left[1+\lambda+e^{\lambda}+{16\over\lambda}\cosh^2\left({\lambda\over2}\right)
\left(\sum_{N=1}^{2n-1}\!'\gamma_N\right)\right], \label{deta1n}
\ee
where 

\be
\gamma_N={1-{2N^2\pi^2\over\lambda^2}\over(4+{N^2\pi^2\over\lambda^2})
(1+{N^2\pi^2\over\lambda^2})^2}.
\ee
Using the second relation of \eqn{prodsum}, we find

\be
\sum_{N=1}^{\infty}\!'\gamma_N={\lambda\over8}\tanh\lambda-{\lambda\over8}\tanh
\left({\lambda\over2}\right)-{\lambda^2\over16}\sech^2\left({\lambda\over2}
\right).
\ee
The function $2e^{\lambda}$ is thus rewritten exactly as
\[{\cosh\lambda\over\cosh^2\left({\lambda\over2}\right)} 
\left\{1+\lambda+e^{\lambda}+{16\over\lambda}\cosh^2\left({\lambda\over2}
\right)\left[{\lambda\over8}\tanh\lambda-{\lambda\over8}\tanh
\left({\lambda\over2}\right)-{\lambda^2\over16}\sech^2\left({\lambda\over2}
\right)\right]\right\},\]
the first factor again expressed as the infinite product of the 
$\alpha_N$s, the expression in the square brackets as the infinite sum 
of the $\gamma_N$s, and approximated by taking the first $n$ terms in the 
product and in the sum.  Note that the zeroth approximation 
$\det A_1^{(0)}=-{1\over2}(1+gFs+e^{gFs})$ 
to \eqn{deta1} is non-perturbative.

The $n$th non-perturbative approximation to 
$-{1\over2}(gF\coth gFs+gF)$ is the ratio of \eqn{deta1n}
and \eqn{detan} (note that the product over the $\alpha_N$s cancel)
is 
and is the object that appears in the $n$th approximation to the Green's function
\eqn{gnth}.  It is not obvious from the reduced form of the matrix $A_1^{(n)}$ that 
the symmetric part of the determinant of the same is the determinant of the 
matrix $\tilde A^{(n)}$.  That is no problem, we find $\det\tilde A^{(n)}$ in 
much the same manner as we found $\det A_1^{(n)}$: take the reduced form of 
$A_1^{(n)}$, reinstate the second column with the operation
col$_2\to$ col$_2+{1\over2}gF$col$_1$, and discard the first row and column to obtain 
the symmetric, reduced form of $\tilde A^{(n)}$:  

\be
\tilde A^{(n)}\sim\left[\matrix{
0&{1\over2}(e^{-gFs}+1)&-{1\over2}gFx_1^T&0&\cdots&-{1\over2}gFx_N^T&0&\cdots\cr
{1\over2}(e^{gFs}+1)&0&0&0&\cdots&0&0&\cdots\cr
{1\over2}gFx_1&0&-\Lone^2&-{3\Lone\over1+\Lone^2}&\cdots&0&0&\cdots\cr
\vspace{-2.0mm}0&0&{3\Lone\over1+\Lone^2}&{5-\Lone^2\over(1+\Lone^2)^2}&\cdots&0&0&\cdots\cr
\vspace{-2.0mm}\vdots&\vdots&\vdots&\vdots&\ddots&\vdots&\vdots&{}\cr
{1\over2}gFx_N&0&0&0&\cdots&-\Ln^2&-{3\Ln\over1+\Ln^2}&\cdots\cr
\vspace{-2.0mm}0&0&0&0&\cdots&{3\Ln\over1+\Ln^2}&{5-\Ln^2\over(1+\Ln^2)^2}&\cdots\cr
\vdots&\vdots&\vdots&\vdots&{}&\vdots&\vdots&\ddots\cr}\right],
\ee 
Expanding the determinant yields

\be
\det\tilde A^{(n)}=-\cosh^2\left({\lambda\over2}\right)
\left(\prod_{N=1}^{2n-1}\!'\alpha_N\right). \label{dettildean}
\ee
Upon comparison of \eqn{dettildean}
and \eqn{deta1n}, noting that $\alpha_N$, $\beta_N$, and
$\gamma_N$ are all even functions of $\lambda$ ($F$), and using 
${1\over2}(1+\cosh\lambda)=\cosh^2\left({\lambda\over2}\right)$, our earlier 
statement that $\det\tilde A^{(n)}$ is the symmetric part of 
$\det A_1^{(n)}$ is apparent.

Finally then, \eqn{deta1n}, the symmetric part thereof,
\eqn{dettildean}, and \eqn{detan} substituted into \eqn{gnth} give the
$n$th approximation to the fermion Green's function in a constant
external field, in the Schwinger-Fock gauge, as defined through the
exact representation \eqn{g1}-\eqn{zeta}.  The approximation easily
leads to Schwinger's exact result \eqn{schwinger} in the limit
$n\to\infty$, and we repeat and emphasise that only the zeroth
approximation corresponds to a perturbative result.

\end{document}